\documentclass[twocolumn,           % Format : preprint, twocolumn
               showpacs,            % Pacs : showpacs, noshowpacs
               preprintnumbers,     % Preprint: preprintnumbers,
               aps,                 % Society: ...
               prd,          	    % Journal Style : pra, prb, prc, prd, pre,
               a4paper,             % Size : a4paper, ...
               groupedaddress,      % Affiliation (Title) : groupedaddress,
               nofootinbib,         % Footnote: footinbib, nofootinbib
               tightenlines,        % Remove additional spaces in a line
               floats,floatfix      % Floating pictures and tables
               ]{revtex4}

\usepackage{graphicx}
\usepackage{latexsym}
\usepackage{amsmath,amssymb}        % amssymb includes amsfonts
\usepackage[draft=false]{hyperref}

\begin{document}

%%--- DRAFTCOPY --------------------------------
%% Prints a large "DRAFT" diagonally across each page
%% Does not show up in TeXview
%% \typeout{Prints "DRAFT" on each page; does not show in TeXView}
%%------------------------------------------------

%======================================%
%<<<<<<<<<<<< TITLE PAGE >>>>>>>>>>>>>>%
%======================================%

\title{Supermassive black holes in scalar field galaxy halos}
\author{L. Arturo Ure\~{n}a-L\'{o}pez}
%\affiliation{Astronomy Centre, University of Sussex, Brighton BN1 9QJ, United 
%Kingdom}
\author{Andrew R.~Liddle}
\affiliation{Astronomy Centre, University of Sussex, Brighton BN1 9QJ, United 
Kingdom}
\date{\today}
\pacs{04.40.-b, 98.35.Jk, 98.62.Gq \hfill astro-ph/0207493}
\preprint{astro-ph/0207493}

%======================================%
%<<<<<<<<<<<<< ABSTRACT >>>>>>>>>>>>>>>%
%======================================%

\begin{abstract}
Ultra-light scalar fields provide an interesting alternative to WIMPS as halo 
dark matter. In this paper we consider the effect of embedding a supermassive 
black hole within such a halo, and estimate the absorption probability and the 
accretion rate of dark matter onto the black hole. We show that the accretion 
rate would be small over the lifetime of a typical halo, and hence that 
supermassive central black holes can coexist with scalar field halos.
\end{abstract}

\maketitle

%======================================%
%<<<<<<<<<<<<<< ARTICLE >>>>>>>>>>>>>>>%
%======================================%

\section{Introduction}
\label{one}

The standard assumption concerning galaxy dark matter halos is that they are 
comprised of some weakly-interacting massive particle (WIMP). However recently 
there has been increased interest in an alternative possibility, that the dark 
matter halo may be comprised of some ultra-light scalar field 
\cite{sin,lee,schunck,peebles,bento,riotto,varun,luis1,siddh,arbey,miguel}. A 
large 
number of such scalar particles, all in their ground state, can be bound by 
their self-gravity; the configurations possess a core radius related to the 
Compton wavelength of the particles in question, and for suitable choices of 
parameters such halos can give a good description of observed rotation curves 
\cite{siddh,arbey}, and optimistically may even provide possibilities to 
alleviate the `cuspy core' \cite{peebles,riotto,arbey,miguel} and `substructure' 
\cite{peebles,varun,luis1} problems of the standard WIMP hypothesis 
\cite{primack,salucci}.

Development of this scenario is at a primitive stage compared to the WIMP 
hypothesis. While it is known that the linear theory evolution of perturbations 
matches the standard scenario, and that time-independent equilibrium 
configurations can broadly reproduce desired halo properties, the scenario has 
not been developed in a full cosmological setting where halo formation is 
tracked. 
Nevertheless, what is known so far is sufficiently intriguing that the scenario 
merits further study.

In this paper, we address one requirement of the scalar-field halo model, which 
is that such a halo must be able to survive the existence of a supermassive 
black hole at its centre, as it is widely believed that such black holes reside 
within many or perhaps even all galaxy halos \cite{laura}. In the WIMP scenario, 
the angular 
momentum of the individual dark matter particles, combined with their low 
interaction rate, ensures that the capture cross-section for halo 
particles by the central black hole is sufficiently small. However, the 
scalar-field halo regime is markedly different; the individual particles do not 
possess angular momentum and indeed are expected to have a Compton wavelength 
upwards of one parsec so that the individual particles occupy a considerable 
volume of space. It is important to verify that the halos are able to survive 
the presence of a central black hole if the scenario is to remain feasible.

This paper is constructed as follows. In Section~\ref{two}, we describe the 
basic steps to 
model a spherical scalar halo without luminous matter. The main intention is to 
provide a simple panorama of the modeling and its appealing properties, such as 
the smooth scalar profiles. In Section~\ref{three}, we use two complementary 
views of the interaction of a scalar halo and a black hole: the classical 
Newtonian picture and the semiclassical approximation. The latter will give us 
information about the absorption probability and the accretion rate of scalar 
matter onto the black hole, the main result of this paper. Finally, we discuss 
the 
main results and some points deserving further investigation.

\section{Scalar field halos}
\label{two}

We briefly describe a galaxy halo assuming that it is made only of 
scalar field matter. A description including, for instance, an exponential 
disk of luminous matter \cite{sin,arbey}, would not significantly change the 
final results. Two similar but distinct kinds of scalar field objects have been 
proposed in the literature to explain galaxy halo structure: boson 
`stars' (comprised of a complex scalar field) \cite{ruffini,mielke,pang,seidel} 
and oscillatons 
(made from a real scalar field) \cite{seidel1,luis,miguel}. For simplicity, we 
restrict 
ourselves to the case of boson stars, though the main results can be easily 
extended to the case of oscillatons. 

The simplest boson stars are those possessing spherical symmetry, for which 
the metric is written in the form
\begin{equation}
ds^2 = g_{\mu \nu} dx^\mu dx^\nu = -B dt^2 + A dr^2 + r^2 \left( d\theta^2 + 
\sin^2 \theta d\varphi^2 \right) \, ,\label{2e1}
\end{equation}
where $A(t,r)$ and $B(t,r)$ are functions to be determined self-consistently 
from the matter distribution. At the classical level, a complex scalar field 
$\Phi$ endowed with a scalar potential $V(|\Phi|)$ is 
described by the energy--momentum tensor
\begin{equation}
T_{\mu \nu} = \frac{1}{2} \left[ \Phi_{,\mu} \Phi^\ast_{,\nu} +  
\Phi^\ast_{,\mu} \Phi_{,\nu} - g_{\mu \nu} \left( \Phi^{,\sigma} 
\Phi^\ast_{,\sigma} + V \right) \right] \,. \label{2e2}
\end{equation}
A self-gravitating boson star is found by solving the coupled 
Einstein--Klein--Gordon (EKG) equations
\begin{eqnarray}
G_{\mu \nu} &=& \kappa_0 T_{\mu \nu} \, , \nonumber \\
\Box \Phi &=& \frac{dV}{d\Phi^\ast} \, , \label{2e3} \\
\Box \Phi^\ast &=& \frac{dV}{d\Phi} \, , \nonumber
\end{eqnarray}
where $G_{\mu \nu}$ is the Einstein tensor corresponding to the 
metric Eq.~(\ref{2e1}), $\kappa_0 = 8\pi G$ (we are taking units such that 
$c=\hbar=1$) and $\Box$ is the covariant d'Alambertian operator.

The EKG equations Eqs.~(\ref{2e3}) admit solutions of the form $\sqrt{\kappa_0} 
\Phi 
= \phi(r) e^{-i\omega t}$. With such an ansatz, the scalar energy--momentum 
tensor Eq.~(\ref{2e2}) and the metric functions $g_{\mu \nu}$ in Eq.~(\ref{2e1}) 
are 
time independent. If we now search for regular and asymptotically flat 
solutions, we should set the boundary conditions $\phi^\prime (r=0)=0, \, 
A(r=0)=1$ and $\phi(r=\infty)=0, \, A(r=\infty)=1$, respectively. The EKG 
equations are then reduced to an eigenvalue problem; for each central value of 
the field $\phi(r=0)\equiv \phi_0$, it is necessary to determine the 
(eigen)values of the fundamental frequency $\omega$ and $B(r=0)\equiv B_0$ to 
find solutions in which the field has $n$ nodes and satisfy the above boundary 
conditions.

In principle, we should also impose the boundary condition $B(r=\infty)=1$. 
However, the eigenvalue problem is further simplified since we can {\it absorb} 
$\omega$ into the metric function $B$. In this way, $\omega$ does not appear 
explicitly in the EKG equations and then it becomes an output value 
determined by $\omega/m =1/\sqrt{B(r=\infty)}$. The normalized temporal metric 
coefficient is calculated via $g_{tt}=-(\omega/m)^2 B(r)$.

According to observations, the gravitational well in galaxies is quite 
weak, which suggests that we should seek boson star solutions in the weak-field 
limit. It is then appropriate to choose a quadratic scalar potential 
$V(|\Phi|)=m^2 |\Phi |^2$ \cite{lee,sin,arbey}. This choice is made not only for 
simplicity, as a quadratic potential can also be considered an approximation to 
more complicated ones possessing a minimum 
\cite{peebles,bento,riotto,varun,luis1,miguel}.

Using the dimensionless radial coordinate $x=r m$, the EKG equations become the 
so-called Schr\"odinger--Newton (SN) equations 
\cite{lee,pang,moroz,arbey,seidel} 
in the weak-field limit: $(\phi_0, \, -g_{tt}-1, \, g_{rr}-1) \ll 1$. Thus, we 
need only solve the simpler set of ordinary differential equations
\begin{eqnarray}
(x\phi)^{\prime \prime} &=& xU\phi \, , \label{2e4} \\
(xU)^{\prime \prime} &=& x\phi^2 \, , \label{2e5}
\end{eqnarray}
where primes denote derivatives with respect to $x$. In order to clarify the 
meaning of function $U(x)$, we take a look at the metric 
coefficients in the weak-field limit,
\begin{eqnarray}
-g_{tt} &\simeq& 1+U(x)-U_\infty \, , \nonumber \\
g_{rr} &\simeq& 1+ xU^\prime (x) \, . \label{2e6}
\end{eqnarray}
Hence, the usual Newtonian potential is given by $U_N = 
(1/2)[U(r)-U_\infty]$, while the value of the fundamental frequency is given by 
$(\omega/m)^{-2} =1+U_\infty$, with $U (x=\infty) \equiv U_\infty$.

Despite its simplicity, the system above still has to be solved numerically, 
with the 
different solutions characterized by, for example, the central 
value $\phi_0$. As in the relativistic case, the solution of 
Eqs.~(\ref{2e4}) and (\ref{2e5}) is an eigenvalue problem; we have to find the 
one 
value $U(0) \equiv U_0$ in order to satisfy the boundary conditions stated above 
and to find $n$-node solutions of the scalar field $\phi(x)$.

To give an order of magnitude estimation of the quantities involved, the 
scalar halo models in the literature \cite{sin,lee,luis1,arbey,miguel} consider 
an ultra-light boson mass $m \sim 10^{-23} \, {\rm eV}$, whose corresponding 
Compton length is $\lambda_C = m^{-1} \sim 1 \, {\rm pc}$. On the other hand, 
the central amplitude of the scalar field would be proportional to the 
gravitational well in galaxies, and then $\phi_0 \sim |U_0| \sim v^2 \sim 
10^{-6}$ with $v$ the rotational velocity of luminous matter in galaxies (in 
units of $c$).

All information of the properties of the scalar halo is contained in 
Eqs.~(\ref{2e4}) and (\ref{2e5}). Of special interest are the asymptotic 
behaviors of 
the scalar and gravitational fields near the centre. It is easy to show that 
\cite{moroz}
\begin{eqnarray}
\phi (x) &=& \phi_0 \left[ 1 + (1/6) U_0 x^2 + {\cal O}(x^4) \right] \, , 
\label{2e7} \\
U(x) &=& U_0 + (1/6) \phi^2_0 x^2 + {\cal O}(x^4) \, , \label{2e8}
\end{eqnarray}
and so the scalar field remains constant up to radii of the order $r \sim  
|U_0|^{-1/2} \lambda_C \sim 1 \, {\rm kpc}$. Therefore, the resulting 
self-gravitating object has a smooth central profile up to distances much larger 
than the Compton length of its particles.

\section{The central black hole}
\label{three}

The geodesics of scalar halos allow massive particles to reach the 
centre of the halo, and in principle the accumulation of matter at the centre is 
not prohibited.\footnote{However, it has been shown that {\it 
rotating} Newtonian boson stars could provide extra repulsive forces at the 
centre \cite{sousa}, which suggests that the inclusion of rotation could avoid 
the excessive accumulation of matter at the centre of scalar objects.} 
Therefore, a black hole can form in the centre of bosonic objects and become a 
threat to their existence. Such a caveat has been recognized before 
\cite{diego}, but it is only recently that the interaction between 
black holes and cosmic scalar fields has begun to be studied seriously 
\cite{jacobson,bean}. 

Our aim now is to outline the physical consequences of the interaction 
between black holes and the scalar halos considered above. For this, we will 
take the two simplest approximations at hand: the classical picture, in which 
the black hole is taken as a central point-like gravitational source, 
and the 
semiclassical picture in which the scalar field lives in the curved space-time 
outside a black hole. As we shall see below, these approximations are 
complementary and can be matched into the scalar halo picture of 
Section~\ref{two}.

\subsection{The classical picture}

Taking into account that the Schwarzschild radius 
\begin{equation}
r_{{\rm s}} \equiv 2GM_{{\rm bh}}\simeq 9.57 \times 10^{-14} {\rm pc} 
\frac{M_{\rm bh}}{M_\odot} 
\end{equation} 
of a central black hole is much smaller than any of the typical 
length scales present in realistic scalar halo (e.g.~the optical radius $r_{{\rm 
opt}}$ or the scalar Compton length $m^{-1}$; see Ref.~\cite{laura} and 
references therein), we can deal with it within the Newtonian regime.

In the classical picture, the total gravitational potential is the superposition 
$U_N =(1/2)[U(r)-U_\infty - r_{\rm s}/r]$, where $U(r)$ is the scalar 
self-gravitational well and $(r_{\rm s}/r)$ is the gravitational field of the 
black hole, which can be seen as a solution in vacuum. Thus, we need only modify 
Eq.~(\ref{2e4}) to include the gravitational influence of the black hole
\begin{equation}
(x\phi)^{\prime \prime} = x \left( U- mr_{{\rm s}}/x \right) \phi \, , 
\label{3e1}
\end{equation}
which resembles the Schr\"odinger equation in a Coulomb-like potential $\sim 
1/r$. We can still construct regular solutions for the scalar field and the 
other metric functions, but we need to change the boundary condition of the 
radial derivative of the scalar field at $r=0$ to
\begin{equation}
\phi^\prime (0) = - \phi_0 m r_{{\rm s}}/2 \, . \label{3e2}
\end{equation}
The other boundary conditions remain the same.

In this classical picture, we notice that the black hole only affects the 
behavior of the field at small $r$, but the scalar profile is still regular. At 
large distances, the scalar profile is unperturbed by the presence of the 
central black hole. In other words, in the Newtonian regime the existence of the 
scalar halo is not threatened by the central gravitational source.

\subsection{The quantum field theory picture}

The approximation we now make is to consider that the scalar field 
lives near the horizon of the black hole in a fixed Schwarzschild background
\begin{equation}
ds^2=-g(r)dt^2 + \frac{dr^2}{g(r)} + r^2 \left( d\theta^2 + \sin^2 \theta 
d\varphi^2 \right) \, , \label{3e3}
\end{equation}
where $g(r)=1-r_{{\rm s}}/r$, and then its properties are determined by the 
field theory in such a curved space-time. This is reasonable since, as stated 
above, the self-gravitating effects of the scalar field appear only at distances 
of 
order $r \gg m^{-1} \gg r_{{\rm s}}$.

Recalling that we are working with a quadratic scalar potential, an $s$-scalar 
wave\footnote{This is the scalar wave with lowest angular momentum $l=0$, and 
hence also the lowest energy. This condition is satisfied for the scalar halos 
considered so far, which are supposed to form a (ground state) Bose condensate. 
The results could be also applied for the case of cosmological scalar fields at 
late times.} obeys the Klein--Gordon equation in metric Eq.~(\ref{3e3})
\begin{equation}
\frac{1}{r^2} \frac{\partial}{\partial r} \left( r^2 g \frac{\partial 
\Phi}{\partial r} \right) - \frac{1}{g} \frac{\partial^2 \Phi}{\partial t^2} = 
m^2 \Phi \, , \label{3e4}
\end{equation}
with the corresponding equation for the complex conjugate field $\Phi^\ast$. 
Eq.~(\ref{3e4}) is separable in the form $\sqrt{\kappa_0} \Phi(t,r)= 
\phi(r)e^{-imt}$, where we have set $\omega = m$, anticipating the classical 
result in which the fundamental frequency does not appear explicitly. The change 
of variable preserves the notation of section~\ref{two}.

At this point, it is convenient to take the Schwarzschild factor $g(r)$ itself 
as the independent variable. Then, the differential equation of $\phi(r)$ near 
the horizon is
\begin{equation}
g^2 \phi^{\prime \prime} + g \phi^\prime + m^2 r^2_{{\rm s}} (1-g)^{-3} \phi 
=0\, , \label{3e5}
\end{equation}
where prime denotes derivative with respect to $g$. 

The ingoing solution of Eq.~(\ref{3e5}) is given, around $g=0 \, (r=r_{\rm 
s})$, in the series form (found using the computer algebra package {\sc maple}, 
{\tt www.maplesoft.com})
\begin{equation}
\Phi(v,r) = \Phi^{(0)} (v,r) \left[ 1- (m r_{{\rm s}})^2 \sum^{\infty}_{n=1} 
\left( P_n 
+imr_{{\rm s}} Q_n \right) g^n \right] \, , \label{3e6}
\end{equation}
where
\begin{equation}
\Phi^{(0)}(v,r)= e^{-im \left[ v-r-r_{{\rm s}}\ln(r/r_{{\rm s}}) \right]} \, . 
\label{3e7}
\end{equation}
Here $v=t+r_\ast$ is the advanced time coordinate defined via the usual 
Kruskal 
coordinate $r_\ast=r+ r_{\rm s} \ln(r/r_{\rm s} -1)$ and we have used the 
relationship
\begin{equation}
g^{r_{\rm s}}(r) = e^{r_\ast - r -r_{\rm s} \ln (r/r_{\rm s})} \, . \label{3e8}
\end{equation}
The coefficients $P_n$, $Q_n$ in Eq.~(\ref{3e6}) have the complicated form
\begin{eqnarray}
P_1 &=& \frac{3}{1+4m^2r^2_{{\rm s}}}, \nonumber \\
P_2 &=& \frac{3}{2^2} \, \frac{2+5m^2r^2_{{\rm s}}+6m^4r^4_{{\rm 
s}}}{(1+m^2r^2_{{\rm s}})(1+4m^2r^2_{{\rm s}})} , \nonumber \\
P_3 &=& \frac{1}{2^2 3^2}\, \frac{40+110m^2r^2_{{\rm s}}+151m^4r^4_{{\rm 
s}}+36m^6r^6_{{\rm s}}}{(1+(4/9)m^2r^2_{{\rm s}})(1+m^2r^2_{{\rm 
s}})(1+4m^2r^2_{{\rm s}})},  \nonumber\\
... &=& ... \nonumber \\
Q_1 &=& \frac{6}{1+4m^2r^2_{{\rm s}}} , \nonumber \\
Q_2 &=& \frac{3}{2^2}\, \frac{2-m^2r^2_{{\rm s}}}{(1+m^2r^2_{{\rm 
s}})(1+4m^2r^2_{{\rm s}})}, \nonumber\\
Q_3 &=& \frac{1}{2^23^3} \, \frac{80-266m^2r^2_{{\rm s}}-319m^4r^4_{{\rm 
s}}-108m^6r^6_{{\rm s}}}{(1+(4/9)m^2r^2_{{\rm s}})(1+m^2r^2_{{\rm 
s}})(1+4m^2r^2_{{\rm s}})}, \nonumber \\
... &=& ... \nonumber
\end{eqnarray} 
In the particular case in which $mr_{{\rm s}} \ll1 $, we can approximate 
$P_n,Q_n$ by their leading terms. Then, we find the approximate expressions
\begin{equation}
P_n \simeq \frac{1}{2n^2} \left( n+1 \right) \left( n+2 \right) \simeq 
\frac{n}{2} Q_n \, . \label{3e9}
%Q_n & \simeq & \frac{1}{n^3} \left( n+1 \right) \left( n+2 \right)
\end{equation}
With these approximate formulas, the sums in Eq.~(\ref{3e6}) can be written in 
terms of known functions, which indicates that the series diverges for $g 
\rightarrow 1 \, (r \rightarrow \infty)$. 

However, we find that for distances $m^{-1} > r \gg r_{{\rm s}}$ (for which we 
can neglect $(r_{\rm s}/r)^2$ and higher-order terms), the radial 
equation for the scalar field becomes
\begin{equation}
\phi^{\prime \prime}+\frac{2}{r} \phi^\prime + m^2 r_{\rm s} \frac{\phi}{r} =0 
\, , \label{3e10}
\end{equation}
where primes now denote derivatives with respect to $r$. The new solutions are 
of the form
\begin{equation}
\phi(r) = r^{-1/2} \left[ C J_1 + D Y_1\right] \left( 2 \sqrt{m^2 r_{{\rm s}} r} 
\right) \, , \label{3e11}
\end{equation}
where $J$ and $Y$ are the Bessel functions of the first and second kind, and $C$ 
and $D$ are 
arbitrary constants. 

The overlap region between the two solutions Eqs.~(\ref{3e6}) and~(\ref{3e11}) 
is $m^{-1} \gg r \gg r_{{\rm s}}$. As we said above, using the approximate 
formulas~(\ref{3e9}), we can estimate the sum of the series in Eq.~(\ref{3e6}). 
For example, if $r=10^3 \, r_{\rm s}$, 
\begin{equation}
\Phi \simeq \Phi^{(0)} (v,10^3 \, r_{\rm s}) \left[ 1- (m r_{{\rm s}})^2 \left( 
512 +14 imr_{{\rm s}} \right) \right] \, .
\end{equation}
The factor $(mr_{\rm s})^2$ highly suppresses the contribution of the series in 
Eq.~(\ref{3e6}), so that we can safely approximate the radial part of the latter 
in this region as
\begin{equation}
\phi(r) \simeq 1-imr^2_{{\rm s}} /r \, . \label{3e12}
\end{equation}

On the other hand, for distances $r \ll m^{-1}$, Eq.~(\ref{3e11}) reduces to
\begin{equation}
\phi(r) \simeq (m^2 r_{{\rm s}})^{1/2} C \left[ 1-\frac{m^2 r_{{\rm s}} r}{2}+ 
\ldots \right] - \frac{(m^2 r_{{\rm s}})^{-1/2} D}{\pi r} \, . \label{3e13}
\end{equation}
Notice that we have included a first-order term in Eq.~(\ref{3e13}), just to 
show that the next-to-order correction is simply the Coulomb-like one, which 
coincides with the classical picture above Eq.~(\ref{3e2}).

Matching Eq.~(\ref{3e13}) onto Eq.~(\ref{3e12}) in the overlap region, we find
\begin{equation}
\frac{D}{C} = i \pi (m r_{{\rm s}})^3 \, , \label{3e14}
\end{equation}
which gives the absorption probability of an $l=0$ spherical wave as \cite{das}
\begin{equation}
\Gamma = 1-\left| \frac{1+\frac{D}{C}e^{i\pi/2}}{1+\frac{D}{C}e^{-i\pi/2}} 
\right|^2 \simeq 4\pi (mr_{{\rm s}})^3 \, , \label{3e15}
\end{equation}
where we have again assumed that $m r_{{\rm s}} \ll 1$.
The interpretation of 
$\Gamma$ is that it gives the fraction of the ingoing wave, and hence the 
fraction of the incoming particles, which is absorbed by the black hole.

The last result indicates that for typical values $mr_{\rm s} \sim 10^{-7}$, 
we have
$\Gamma \sim 10^{-20}$ which implies that the absorption of the scalar field is 
negligible and that, from the semi-classical point of view too, a central black 
hole and a scalar halo can be put together. Eq.~(\ref{3e15}) coincides with 
previous calculations, which also indicate that the absorption probability of 
higher $l$-modes is further suppressed by a factor of the order $(m r_{\rm 
s})^{2l}$ \cite{page}. 

Summarizing, we can say that the solutions of the scalar halo are given by 
Eqs.~(\ref{3e1}) and (\ref{2e5}) for $r \geq m^{-1}$, by Eq.~(\ref{3e11}) for 
$m^{-1} > r \gg r_{{\rm s}}$ and by Eq.~(\ref{3e6}) for $r \sim r_{{\rm s}}$, 
with the 
absorption probability Eq.~(\ref{3e15}) calculated in the overlap region 
$r_{{\rm s}} \ll r < m^{-1}$. Formally speaking, the three different solutions 
are well matched to each other if we multiply Eqs.~(\ref{3e6}) and (\ref{3e11}) 
by the 
central amplitude $\phi_0$ calculated for the scalar halo in 
Eqs.~(\ref{3e1}) and (\ref{2e5}). Since the latter is just an overall factor, 
the 
absorption probability Eq.~(\ref{3e15}) remains the same.

Observe that the second solution in Eq.~(\ref{3e11}) could have been obtained 
within the classical picture in Eq.~(\ref{3e1}), but it was not taken into 
account because it diverges at the origin and our purpose was to construct 
regular solutions. But, as we have seen in this section, this second solution 
contains the information of the interaction between the black hole and the 
scalar field, since it is through it that we obtained a non-null absorption 
probability.

Another important issue that can be calculated is the accretion rate of the 
scalar field into the black hole by using the formula Eq.~(3.1) in 
Ref.~\cite{jacobson}. The scalar energy--momentum tensor should be written in 
the new variables $(v,r)$, and then we obtain for the flux of Killing energy 
across the horizon
\begin{equation}
dM/dt = 4\pi r^2_{\rm s} \times T_{vv}(v,r_{\rm s}) = (2G)^{-1} (\phi_0 m 
r_{{\rm s}})^2 \, , \label{3e16}
%T_{vv}(v,r_{\rm s}) &=& \Phi_{,v} \Phi^\ast_{,v} \, , \nonumber
\end{equation}
in which we have included the overall factor $\phi_0$. Using typical numbers 
$\phi_0 mr_{\rm s} \sim 10^{-13}$, the accretion rate is 
quite small, $dM/dt \simeq 10^{-14} M_\odot {\rm y}^{-1}$, a result that is 
consistent with the small absorption probability given by Eq.~(\ref{3e15}).

\section{Conclusions}
\label{four}

We have analyzed the impact of a central supermassive black hole on galactic 
halos comprised of ultra-light scalar particles. From simple physical grounds, 
we should expect that the accretion rate of a scalar halo onto a black hole is 
small, since the boson particles cannot `fit' into the horizon due to their 
large Compton length. Besides, the absorption probability should be proportional 
to the ratio of the effective `area' of the two objects, and hence proportional 
to  $(mr_{\rm s})^2$ as happens for massless scalar fields \cite{das}.

We found that the absorption probability is decreased by an extra factor 
$mr_{\rm s}$, which assures the coexistence of the bosonic halo and the 
central black hole. For this, we showed how to construct consistent and regular 
solutions on different scales. In addition, the accretion rate of scalar matter 
onto the black hole is so small that the matter absorbed by the black hole is 
much less than a solar mass in the whole lifetime of the Universe. On the other 
hand, this result would indicate that the current observed accretion in galaxy 
black holes would not be due to matter provided by a scalar halo. 

We have only investigated the equilibrium state of a relaxed scalar halo and 
a central black hole, and it would be interesting to have a more dynamical view 
studying the formation (simultaneously or not) of the two 
objects. This would require of the evolution of the full Einstein 
equations, which is well beyond the scope of this paper.

A related issue is the interaction of primordial black holes with cosmic 
scalar fields, prior to the gravitational collapse of density perturbations, as 
recently outlined in Refs.~\cite{jacobson,bean}. For a cosmic scalar 
field endowed with a quadratic potential, the accretion rate would also be given 
by formula Eq.~(\ref{3e16}), and then the field would have the oscillatory 
behavior Eq.~(\ref{3e6}) near the black hole horizon. That is, the mass of the 
bosonic field still prevents a strong interaction between black holes and cosmic 
scalar fields. Other kind of scalar potentials would lead to more interesting 
pictures \cite{bean}.

%======================================%
%<<<<<<<<<<< ACKNOWLEDGEMENTS >>>>>>>>>>%
%======================================%

\begin{acknowledgments}
L.A.U.-L.~was supported by CONACyT, M\'exico under grant 010385, and A.R.L.~in 
part by the Leverhulme Trust. We thank Ricardo Becerril, F. Siddhartha Guzm\'an, 
Julien Lesgourgues, Tonatiuh Matos, Lu\'{\i}s Mendes and Ian Moss for useful 
discussions.
\end{acknowledgments}

%======================================%
%<<<<<<<<<<<< BIBLIOGRAPHY >>>>>>>>>>>>%
%======================================%

%%%%%%%%%%%%%%%%%%%%%%%%%%%%%%%%%%%%%%%%%%%%%%%%%%%%%%%%%%%%%%%%%%%%%%%%

\begin{thebibliography}{}
\bibitem{sin} S. J. Sin, Phys. Rev. D {\bf 50}, 3650 (1994), {\tt
	hep-ph/9205208}; S. U. Ji and S. J. Sin, Phys. Rev. D {\bf 50}, 
	3655 (1994), {\tt hep-ph/9409267}.
\bibitem{lee} J. W. Lee and I. G. Koh, Phys. Rev. D {\bf 53}, 2236 (1996), 
	{\tt hep-ph/9507385}.
\bibitem{schunck} F. E. Schunck, astro-ph/9802258.
\bibitem{siddh} F. S. Guzm\'an and T. Matos, Class. Quantum Grav. {\bf 17}, L9
	(2000), {\tt gr-qc/9810028}; T. Matos, F. S. Guzm\'an, and 
	D. N\'u\~nez, Phys. Rev. D {\bf 62}, 061301 (2000), {\tt
	astro-ph/0003398}; T. Matos and F. S. Guzm\'an, Class. Quantum 
	Grav. {\bf 18}, 5055 (2001), {\tt gr-qc/0108027}.
\bibitem{peebles} P. J. E. Peebles, Astrophys. J. {\bf 534}, L127 (2000), {\tt
	astro-ph/0002495}; J. Goodman, New Astron {\bf 5}, 103 (2000), {\tt
	astro-ph/0003018}.
\bibitem{bento} M. C. Bento, O. Bertolami, R. Rosenfeld, and L. Teodoro, Phys.
	Rev. D {\bf 62}, 041302 (2000), {\tt astro-ph/0003350}; O. Bertolami, 
	M. C. Bento, and R. Rosenfeld, astro-ph/0111415.
\bibitem{riotto} A. Riotto and I. Tkachev, Phys. Lett. B {\bf 484}, 177 (2000),
	{\tt astro-ph/0003388}.
\bibitem{varun} V. Sahni and L. Wang, Phys. Rev. D {\bf 62}, 103517 (2000),
	{\tt astro-ph/9910097}.
\bibitem{luis1} T. Matos and L. A. Ure\~na-L\'opez, Class. Quantum Grav.
	{\bf 17}, L75 (2000), {\tt astro-ph/0004332}, Phys. Rev. D {\bf 63},
	063506 (2001), {\tt astro-ph/0006024}, Phys. Lett. B {\bf 538}, 246
	(2002), {\tt astro-ph/0010226}; J. E. Lidsey, T. Matos, and L. A.
	Ure\~na-L\'opez, Phys. Rev. D {\bf 66}, 023514 (2002),                         {\tt astro-ph/0111292}.
\bibitem{arbey} A. Arbey, J. Lesgourgues, and P. Salati, Phys. Rev. D {\bf 64},
	123528 (2001), {\tt astro-ph/0105564}; {\it ibid}, {\bf 65}, 083514
	(2002), {\tt astro-ph/0112324}.
\bibitem{miguel} M. Alcubierre, F. S. Guzm\'an, T. Matos, D. N\'u\~nez, 
	L. A. Ure\~na-L\'opez, and P. Wiederhold, Class. Quantum Grav.                 {\bf 19}, 5017 (2002), {\tt gr-qc/0110102}; {\tt astro-ph/0204307}.
\bibitem{primack} J. R. Primack, {\tt astro-ph/0112255}, 
	{\tt astro-ph/0205391}.
\bibitem{salucci}  P. Salucci, F. Walter, and A. Borriello, {\tt 
	astro-ph/0206304}.
\bibitem{laura} L. Ferrarese, {\tt astro-ph/0203047}, {\tt astro-ph/0207050}.
\bibitem{ruffini} R. Ruffini and S. Bonazzola, Phys. Rev. {\bf 187}, 1767
	(1969).
\bibitem{pang} R. Friedberg, T. D. Lee, and Y. Pang, Phys. Rev. D {\bf 35}, 
	3640 (1987).
\bibitem{seidel} E. Seidel and W-M. Suen, Phys. Rev. D {\bf 42}, 384 (1990).
\bibitem{mielke} E. W. Mielke and F. E. Schunck, {\it Proceedings of the 8th
	Marcel Grossmann Meeting}, Jerusalem, Israel (World Scientific,
	Singapore, 1999), {\tt gr-qc/9801063}.
\bibitem{seidel1} E. Seidel and W-M. Suen, Phys. Rev. Lett. {\bf 66}, 1659
	(1991).
\bibitem{luis} L. A. Ure\~na-L\'opez, Class. Quantum Grav. {\bf 19}, 2617
	(2002), {\tt gr-qc/0104093}; L. A. Ure\~na-L\'opez, T. Matos, and 
	R. Becerril, in	preparation.
\bibitem{moroz} I. M. Moroz, R. Penrose, and P. Tod, Class. Quantum Grav.
	{\bf 15}, 2733 (1998); P. Tod and I. M. Moroz, Nonlinearity {\bf 12},
	201 (1999).
\bibitem{sousa} V. Silveira and C. M. G. de Sousa, Phys. Rev. D {\bf 52}, 
	5724 (1995), {\tt astro-ph/9508034}.
\bibitem{diego} D. F. Torres, S. Capozziello, and G. Lambiase, Phys. Rev. D
	{\bf 62}, 104012 (2000), {\tt astro-ph/0004064}; D. F. Torres, Nucl.
	Phys. B {\bf 626}, 377 (2002), {\tt hep-ph/0201154}.
\bibitem{jacobson} T. Jacobson, Phys. Rev. Lett. {\bf 83}, 2699 (1999),
	{\tt astro-ph/9905303}.
\bibitem{bean} R. Bean and J. Magueijo, {\tt astro-ph/0204486}.
\bibitem{das} S. R. Das, G. Gibbons, and S. D. Mathur, Phys. Rev. Lett. 
	{\bf 78}, 417 (1997), {\tt hep-th/9609052}.
\bibitem{page} D. N. Page, Phys. Rev. D {\bf 13}, 198 (1976); W. G. Unruh,       
      Phys. Rev. D {\bf 14}, 3251 (1976).
\end{thebibliography}
\end{document}